\documentclass{aa}  

\usepackage{graphicx}
\usepackage[labelfont = bf]{subcaption}
\usepackage{upgreek}
\usepackage{xcolor}
\usepackage{lscape}
\usepackage{svg}
\usepackage{txfonts}
\usepackage[breaklinks=true]{hyperref}
\usepackage{orcidlink}
\begin{document}

   \title{ApPolLO: \textbf{Ap}ertif counterparts of \textbf{Pol}arised \textbf{LO}FAR sources}


   \author{A. Berger\inst{1,2} \corrauth{aberger@physik.uni-bielefeld.de} \orcidlink{https://orcid.org/0000-0001-5022-2785}
   \and B. Adebahr\inst{2} \email{adebahr@astro.rub.de} \orcidlink{https://orcid.org/0000-0002-5447-6878}
   \and H. Dénes\inst{3} \email{helgadenes@gmail.com} \orcidlink{https://orcid.org/0000-0002-9214-8613}
   \and R.-J. Dettmar\inst{2} \email{dettmar@astro.rub.de} \orcidlink{https://orcid.org/0000-0001-8206-5956}
   }

   \institute{Universität Bielefeld, Faculty of Physics, Postfach 100131, 33501 Bielefeld, Germany
   \and Ruhr University Bochum, Faculty of Physics and Astronomy, Astronomical Institute (AIRUB), Universitätsstrasse 150, 44780 Bochum, Germany
   \and College of Sciences and Engineering, Universidad San Francisco de Quito, Quito, 170901, Ecuador
   }

   \date{}

 
  \abstract
   {Knowledge of the structure and evolution of cosmic magnetic fields is indispensable to understand how primordial magnetic fields were generated and amplified during the lifetime of the Universe. Polarisation studies over significant redshift ranges are an important tool in understanding cosmic magnetic fields.}
   {We aim to understand the structure and evolution of cosmic magnetic fields, utilising the polarised spectra and depolarisation characteristics of (background) radio sources.}
   {Using radio polarisation data at two different frequencies from the LOFAR Rotation Measure (RM) grid catalogue at 144\,MHz and new Apertif data at 1.4\,GHz, we investigate the source statistics of a common sample of polarised sources, deriving the depolarisation ratio between these two frequencies. We utilise this to disentangle depolarising effects local to the source from those along the line of sight.}
   {We compile a catalogue of 598 polarised radio components, most of which are located in low-density environments. Fractional polarisation decrease for both wavelengths with redshift. We attributed the main depolarisation at 1.4\,GHz to effects occurring close to the sources, while the depolarisation measured at 144\,MHz is most likely generated by cosmic filaments along the line-of-sight. We were able to get a first redshift dependent upper limit estimate of the RM dispersion in cosmic web filaments.}
   {}


   \maketitle
   \nolinenumbers
%

\section{Introduction}

The genesis and evolution of magnetic fields in the Universe is an open question. While some theories suggest primordial magnetic fields were created during the Big Bang, other theories assume magnetic fields to be formed by physical processes right after the Big Bang \citep{Widrow2002}. A critical and still unsolved question within these theories is the amplification of such weak primordial magnetic fields to the $\upmu$G field strength we observe in the local Universe nowadays \cite[e.g.][and references therein]{Akahori2011,OSullivan2020, Carretti2025}.

The most dense regions in the Universe, clusters and super-clusters are connected by less dense filaments surrounding voids and forming a web-like structure, the so-called cosmic web \citep{Nelson2019}.
While magnetic fields in clusters are strongly altered, masking their origin, baryonic matter and magnetic fields in filaments are less processed, making them an ideal target for research on primordial conditions \citep{Vazza2021a}.

The emission from those filaments is extremely faint, making direct detections exceedingly challenging. 
However, \cite{Vernstrom2021} showed that via stacking it is possible to detect those filaments at low radio frequencies. Besides direct emission, we can also trace the imprint of cosmic web filaments on the polarised spectra of background sources due to the effect of Faraday Rotation, causing the polarisation angle $\psi$ to change as a function of wavelength $\lambda$ and the so called Rotation Measure (RM) as 
\begin{align}
   \Delta \psi &= RM \lambda^2\\
   RM &= 0.812 \int \left(\frac{n_e}{cm^{-3}}\right) \left(\frac{B_{||}}{\mu G}\right) \left(\frac{dr}{pc}\right).
   \label{eq:RM}
\end{align}
Here $n_e$ is the electron density, $B_{||}$ the magnetic field strength component parallel to the line of sight and $dr$ is the path length.

In the case of turbulence in the magnetic fields along the line of sight, this effect causes the signal to depolarise.
The depolarisation that happens somewhere between source and observer is called external Faraday Dispersion and can at first order be described by the Burn-law \citep{Burn1966,Sokoloff1998}:
\begin{align}
    FP &= FP_0 \exp(-2 \sigma_{RM}^2 \lambda^4)
    \label{eq:extrenal_depol}\\
    \text{with } \sigma_{RM}^2 &= 0.81 \frac
    {\left( \langle n_e\rangle \langle B_{turb}\rangle \right)^2 L \,d}{f},
    \label{eq:sigma_square}
\end{align}
where $FP$ being the measured and $FP_0$ the emitted fractional polarisation, $\lambda$ the wavelength and $\sigma_{RM}$ the dispersion of the Rotation Measure, which is dependent on the average electron density $\langle n_e\rangle$ and the  average strength of the turbulent magnetic field $\langle B_{turb}\rangle$ over the path-length $L$. $d$ is the correlation length of the turbulent cells in pc and $f$ the filling factor.

Investigating the magnetic fields in cosmic filaments through their impact on the polarised background signal requires a statistical approach and therefore large samples of background sources. Using 1.4\,GHz Westerbork Synthesis Radio Telescope (WSRT) data from the Lockman Hole \cite{Berger2021} found an anti-correlation between redshift and fractional polarisation already in a smaller sample of 150 polarised radio sources. This trend was interpreted to originate from an evolution of either the sources' direct environment or the magnetic fields in the cosmic web.

The LOw Frequency ARray (LOFAR) Two metre Sky Survey \citep[LoTSS,][]{Shimwell2022} DR2 RM grid catalogue from \cite{OSullivan2023} provides a unique low-frequency sample of polarised sources at 144\,MHz. This sample was already utilised by \cite{Carretti2022} in combination with NVSS 1.4\,GHz data. The authors confirmed a redshift dependence of the polarised signal and could even attribute the 1.4\,GHz signal to the environment local to the sources and the 144\,MHz signal to a signature by cosmic web filaments. In the above mentioned and successive studies \citep{Carretti2023,Carretti2025} they could also constrain the magnetic field strength of the cosmic filaments to several tenths of nG while also highlighting the significance of foreground corrections.

\citet{Akahori2011} used model derived from cosmological structure formation simulations to examine the magnitude of the $RM$ with redshift distance. They found an increase up to z=1. \citet{Mtchedlidze2024} used cosmological magnetohydrodynamic simulations for evolving primordial magnetic field strengths during large-scale structure formation to generate a sample of mock $RM_{IGM}$ data. Both aforementioned studies are in agreement with the results by \cite{Carretti2023} and \citet{Carretti2025}. 

As already illustrated by \cite{Carretti2022} the strong wavelength dependency of depolarisation effects helps to disentangle between the different depolarising components, the one local to the source and those caused by filaments along the line of sight. In this work, we therefore combine data from the LoTSS DR2 RM grid catalogue with new data from the APERture Tile In Focus \cite[Apertif,][]{VanCappellen2022}
to investigate the depolarisation of extragalactic sources over cosmic time.



This paper is organised as follows, Sec. \ref{sec:Data} describes the data used and the data reduction and imaging of the Apertif data. The Data analysis and catalogue creation is presented in Sec. \ref{sec:data_analysis}. We will give an overview of the catalogue in Sec. \ref{sec:overview}. The redshift dependent depolarisation will be discussed in Sec. \ref{sec:depol} and the depolarisation from cosmic filaments in Sec. \ref{sec:depol_filaments}. In Sec. \ref{sec:discussion} we will discuss the results and a summary is given in Sec. \ref{sec:summary}.

\section{Data Reduction}
\label{sec:Data}

We used the LOFAR DR2 RM grid catalogue from \cite{OSullivan2023} as a starting point of our investigation and identified those sources, where reduced Apertif data was already available. 
Apertif observed the northern sky at 1.4\,GHz in a two-tier survey \citep{Adams2019}: 
the Apertif Wide Extragalactic Survey (AWES) and the Apertif Medium Deep Extragalactic Survey (AMES). 
Apertif observations took place between 2020 and 2022, with each single observation having an observation time of 11.5\,h.
The data reduction of both surveys was performed at Astron using the Apertif calibration pipeline \cite[\texttt{Apercal},][]{Adebahr2022a}.
Even though Apertif stopped operation early 2022, the data reduction was not completely finished by the time this work started. We used data that was reduced until September 2022. An investigation using the full Apertif footprint, covering an area of $\sim 2200\,$deg$^2$ will be presented in a follow-up paper.

Using a list of the reduced data, we identified the Apertif fields and beams that contribute to the position of the LOFAR RM grid sources, with a primary beam response of at least 10\,\%. 
In total, we found 731 sources with available Apertif data. 

For the majority, namely 595 IDs, each contributing beam was only observed once, while for the other 136 IDs, at least one contributing beam was observed multiple times.
While some AWES fields were observed twice, to increase data quality, compensate for RFI and aim for a deeper survey, AMES fields were observed up to 10 times.
The combination of Apertif polarisation data of multiple observations for one beam is currently under investigation and therefore not yet solved.
Therefore, we used the metadata (resolution information and noise statistics) of the beams in question for those 136 IDs, to identify the observation with the best quality to be used for our analysis.

For the imaging a custom pipleline was utilised, 
that first adjusts the number of subbands used \citep[see][for details]{Berger2025}, and then runs \texttt{amosaic} and \texttt{aperpol} \citep{Adebahr2022b} on the data for each individual ID.
The adjustment in the number of subbands was necessary, since the observational frequency of Apertif was shifted in January 2021 due to RFI covering a large part of the lower observing frequencies of Apertif \citep{Adams2022}.

For 132 IDs the imaging was not possible, this was mainly the case if no beam passed the constraints for data quality\footnote{We set constraints via the configuration file for \texttt{amosaic} to be a maximum RMS noise of 35$\,\upmu$Jy, and a major and minor beam resolution of 20.5\arcsec (28\arcsec) and 16\arcsec (16\arcsec) for total (polarised) intensity.}. For 8 of those datasets we had total intensity but no polarisation data available. 

The typical resolution of Apertif observations is 15\arcsec $\times$ 15\arcsec$\cdot \cos{\text{DEC}}$. Leading to a typical resolution of 15\arcsec $\times$ 18\arcsec for our sample. 

\section{Data analysis}
\label{sec:data_analysis}

\texttt{Aperpol} creates a polarised intensity mosaic from the individual compound beams as well as a source catalogue of the entire mosaic, using pyBDSF\citep{Mohan2015}.
Each polarised source in each of those catalogues was then automatically crossmatched with the Apertif total power (TP) catalogue as generated from the total intensity map generated from the same data, as well as with ancillary data from the NRAO VLA Sky Survey \citep[NVSS,][]{Condon1998}, the All-source catalogue from the Wide-field Infrared Survey Explorer satellite \cite[AllWISE,][]{Cutri2021} and the Sloan Digital Sky Survey \cite[SDSS,][]{Ahumada2020} \cite[see][for details]{Adebahr2022a}.
This type of catalogue was generated for the mosaics around 599 of the LOFAR DR2 RM grid sources.

\subsection{Cross-match}

To create a catalogue of sources with Apertif and LOFAR information (ApPolLo sources hereafter), we started with the 599 LOFAR RM grid catalogue sources for which we have now imaged their Apertif counterparts.
In order to identify these Apertif counterparts we cross-matched each LOFAR RM grid catalogue source position with the corresponding Apertif catalogue of the imaged field and took the closest Apertif source within 60$\arcsec$ of this position, as the preliminary counterpart.
The search radius was chosen to account for the RM grid catalogue using the highest signal to noise pixel as the position, while for the Apertif catalogue considers the full extent of the source, as it is based on pyBDSF.
For the majority of sources the polarised emission at 1.4\,GHz is more extended than at 144\,MHz, which also made automatic cross-matching more complicated. 
We therefore decided to start with a rather high search radius and confirm the preliminary cross-matches manually to guarantee reliability.
This step produced a cross-match catalogue with 599 rows that contains the LOFAR RM grid catalogue information as well as the information from the cross-matched Apertif source. In cases where no Apertif source was within range, we added empty rows.

To verify this cross-match, we generated plots, similar to those created by \texttt{aperpol}\citep[see Figure 4 of][]{Adebahr2022a}, but with information on the LOFAR 144\,MHz data (LOFAR $PI$ and $TP$ maps with corresponding source positions marked and a LOFAR FP map with contures of the LOFAR PI emission) exchanging them with three of the four WISE images. 
One plot was generated per source ID in the cross-match table, including those without an Apertif counterpart. 

Following manual verification and, where necessary, adjustment of the LOFAR-Apertif cross-matches, we also inspected the automatic cross-matches with Apertif TP, NVSS, WISE and SDSS performed by \texttt{aperpol} for ApPolLo sources. 
To this end we collected the \texttt{aperpol} output plots of the cross-matched Apertif IDs for quick visual inspection and only interactively adjusted the catalogues for these sources using the \texttt{aperpol} interactive module.
We then redid the plots, ran the cross-match again, this time considering the interactive catalogues, collected the plots and started another round of interactively correcting the catalogues. This was repeated until no further interaction was needed.

For some sources, even after manual interaction, no Apertif cross-match was found. This was mainly the case for large extended sources with diffuse emission where pyBDSF could not properly identify the source. 
However, this value is only used to compare the ApPolLo sample to an Apertif reference sample in Sec. \ref{sec:compare_apertif_ref}, we therefore simply exclude those from the analysis in Sec. \ref{sec:compare_apertif_ref}.

\subsection{Peak finding}
\label{sec:peak_finding}

Since the Apertif catalogues contain information on the integrated flux and no accurate peak flux information \citep[see][]{Berger2025}, while the LOFAR RM grid catalogue only contains peak flux information, these values are not directly comparable. For this reason, we also added the flux and $RM$ information from the Apertif image at the exact pixel position of the LOFAR RM grid source position to the catalogue.

The peak of the emission at 1.4\,GHz does not need to be at the same position as the peak at 144\,MHz. In fact, for most of our sources these two peaks are not aligned.
Therefore, we identified and measured the peak of the 1.4 GHz component that corresponds to the relevant LOFAR RM grid source position.

Source finding in polarised intensity remains an open problem with several different approaches possible.
Simply following the exact same approach as used for the RM grid catalogue is not practical here, since in general due to depolarisation effects sources have a much higher fractional polarisation at L-Band compared to 144\,MHz \citep{Sokoloff1998,OSullivan2023}.

To identify the peak position (or in case of multi-component sources the peak positions) of our sources, we developed a scheme based on the one of \citet{OSullivan2023}: 
A noise map ($\sigma_{QU}$ was generated as by using the median absolute deviation from the median (MAD) along the Faraday axis for every pixel.
Using the noise map and the PI-map, all pixels with a signal-to-noise ratio $<7\,\sigma_{QU}$ were blanked.
Polarised emission is only expected at the location of sources, therefore the total intensity mask, as generated by pyBDSF when running \texttt{aperpol}, is also applied to generate the 'clipped-masked-map'.
This masking step significantly reduced the contamination by noise pixels. 
However, to avoid counting a single pixel exceeding the $7\,\sigma_{QU}$ threshold as a polarised source, we utilised the \textsc{scipy.ndimage.lable}, \textsc{generate\_binary\_structure} and \textsc{find\_objects} functions to filter for islands that contain at least six connected pixels.

This last step follows the procedure used by pyBDSF. Six pixels is also the minimum island size pyBDSF would use per default based on our grid and resolution. 
We then generated a peak\_catalogue by simply identifying the pixel with the maximum flux within each island. 

However, this method is only sensitive to the main peak within a source, so minor peaks in multi component sources will not be detected.
To overcome this, we also utilised the \textsc{findpeaks}\footnote{\url{https://erdogant.github.io/findpeaks/pages/html/Topology.html}} package's \textsc{findpeaks} task with the 'topology' method on our clipped-masked-map, aiming to identify all peaks within an island.
We call the catalogue generated in this way the peak\_finder\_catalogue.

Following a similar procedure as before, for every source in our cross-matched catalogue, the peak in the respective peak\_finder\_catalogue, closest to the position of the LOFAR RM grid source was identified within 20\arcsec and added to the cross-matched catalogue.
For visual inspection, we once again adjusted the plots; all images were regridded to the Apertif $PI$ map, and the plot now also hosts the clipped-masked-map and the corresponding SNR map, including markers for the peak positions.
We used those plots to manually adjust the cross-matches.
In some cases, by visual inspection, the threshold for clipping was reduced to 6 or even $5\,\sigma$.
Even though a lower threshold increases the number of false positive detections, this is not crucial in the case of ApPolLo sources, since the sources are already known to be polarised at 144\,MHz and, due to the order of magnitude difference and the strong depolarisation effects at low frequencies, we assumed that polarised emission for all of our sources is also detectable at L-Band.
However, for two sources, no Apertif counterpart was found above a threshold of $5\,\sigma_{QU}$. These two sources will be discussed in detail in Sec. \ref{sec:non-detect}.


During the described close inspection of our generated plots, we found that IDs 549 and 20245 appear to be the same source, with all Apertif peak positions being closer to the position of ID 549 than ID 20245. For this reason, we decided to exclude ID 20245 from the analysis, reducing the full catalogue to 598 sources.

After identifying the peak and its position for each source, we also added the $RM$, $TP$, $FP$ and $SNR$ value for this position to the catalogue. 


\section{Data overview}
\label{sec:overview}

The resulting catalogue contains 598 sources with information on the sources at 1.4\,GHz and 144\,MHz. 
Out of these, 
two do not show polarised intensity in our Apertif maps above $5\,\sigma$.
Even though we have simply assumed all LoTSS DR2 RM grid IDs to be individual sources so far, according to the identification in the LoTSS DR2 value added catalogue \citep{Hardcastle2023}, those are actually 598 components, belonging to 560 different sources. 
Since we did not apply any selection criteria to our sample other than Apertif data to be available, we assume it to be an unbiased subset of the catalogue of \cite{OSullivan2023}. 

\subsection{Comparison to the Apertif sample}
\label{sec:compare_apertif_ref}

Previous studies \citep[e.g.][]{OSullivan2020,Stuardi2020,Carretti2022} found sources with detectable polarised emission at 144\,MHz tending to be located in low density environments, leading to a strong selection effect when comparing the ApPolLo sample to all Apertif polarised sources.
Since a complete polarised intensity map and catalogue of AWES is still work-in-progress,
we used the combined GOODS-N \citep[1182 polarised sources in an area of 47.4\,$deg^2$,][]{Berger2025} and SVC \citep[1170 polarised sources in an area of 56.4$\,deg^2$,][]{Adebahr2022b} samples as a reference for the entire survey.
For a consistent comparison with the Apertif sample we need to use the integrated fluxes as obtained by \texttt{aperpol} ($PI_{Isl}$), while for comparison with the LOFAR RM grid catalogue values we use either the Apertif peak flux ($PI_{Peak}$) or the Apertif flux at the 144\,MHz peak position ($PI_{pos}$). 
This is to address the LoTSS DR2 RM Grid catalogue containing only single-pixel peak fluxes, which cannot be properly compared to integrated fluxes. 

\subsubsection{Fractional Polarisation PDF}
\label{sec:FP_PDF}
The Apertif GOODS-N data show a median fractional polarisation of $5.68\,\%$ when accounting for the complete sample and $4.48\,\%$ when excluding sources with $TP \leq 10\,$mJy, where the sensitivity limitation \citep[see][]{Berger2025} has the strongest effect. 
The low-density environment, favoured by polarised sources found in low frequency observations, results in less depolarisation even at cm wavelengths. Therefore, the higher median fractional polarisation of the ApPolLo sources at 1.4\,GHz of $FP_{Isl}^{Med} = 7.44\,\%$ ($7.37\,\%$ for $TP \leq 10\,$mJy) aligns with our expectations. 

\begin{figure}[hbtp]
    \centering
    \includegraphics[width = \linewidth]{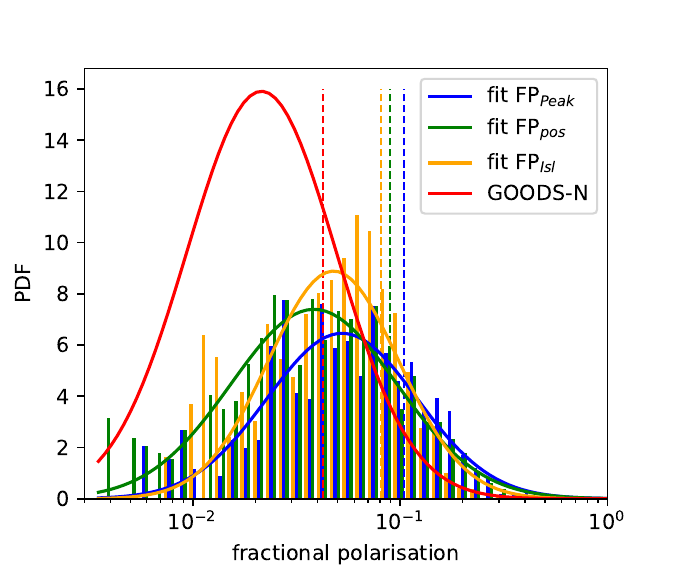}
    \caption[Fractional polarisation PDF for ApPolLo sources]{PDF of the fractional polarisation at 1.4\,GHz of the ApPolLo sources. The orange data gives $FP_{Isl}$ as calculated by \texttt{aperpol}, blue uses the fractional polarisation taken from the peak position of the Apertif polarised component, and green data uses the Apertif flux at the exact position of the LOFAR source. The red line is the best fit PDF from the Apertif GOODS-N data from \cite{Berger2025}. The thin vertical lines indicate the mean value of the underlying normal distribution resulting from the fit.}
    \label{fig:ap:FP_PDF}
\end{figure}

In Fig. \ref{fig:ap:FP_PDF}, we show the PDF for the different fractional polarisation values calculated for the ApPolLo sample compared to the PDF of the GOODS-N sample. The fits shown are using the same log-normal distribution as defined in Eq. 7 of \cite{Berger2025}:
\begin{align}
    f(FP) &= \frac{log_{10}(e)}{FP \sigma \sqrt{2\pi}} \exp{-\frac{(log_{10}(FP) - \mu)^2}{2\sigma^2}},
    \label{eq:PDF}
\end{align}
The best fit parameters are provided in Table \ref{tab:PDF_fit}.

\begin{table}[]
    \centering
    \caption{Best fit values for eq. \ref{eq:PDF}}
    \begin{tabular}{c|c|c}
         & $\mu$ & $\sigma$ \\
         \hline
         \hline
        $FP_{Peak}$ & $-0.979 \pm 0.019$ & $0.362 \pm 0.015$ \\
        $FP_{pos}$ & $-1.048 \pm 0.017$ & $0.400 \pm 0.013$ \\
        $FP_{Isl}$ & $-1.09 \pm 0.02$ & $0.314 \pm 0.016$\\
    \end{tabular}
    \label{tab:PDF_fit}
\end{table}

\subsubsection{Flux Density}
\begin{figure}
    \centering
    \includegraphics[width = \linewidth]{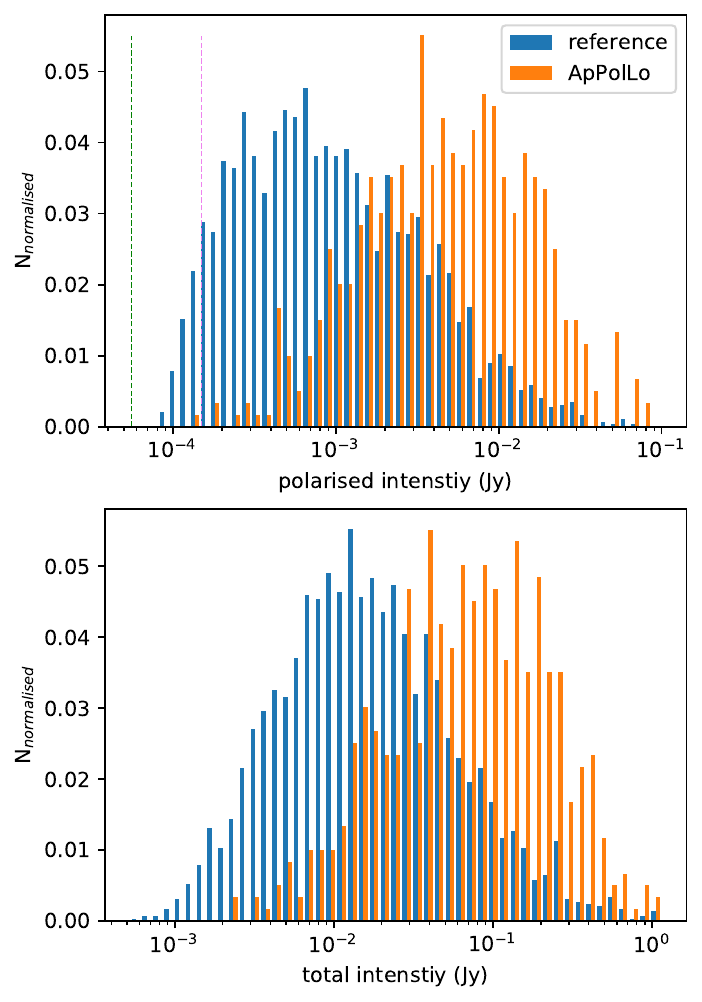}
    \caption[Flux histogram of the ApPolLo and Apertif reference samples]{Histogram of the polarised (upper panel) and total (lower panel) intensity of the ApPolLo sample, compared to the Apertif reference sample (GOODS-N and SVC). The vertical dashed lines in the upper panel show the flux measured at the position of the two sources 340 (violet) and 411 (green) with no peak intensity found above $5\sigma_{QU}$.}
    \label{fig:ap:PI_TP_hist}
\end{figure}

$~$\\
We find that the ApPolLo sources have a significantly higher median total and polarised intensity flux density of $TP = 70.47\,$mJy and $PI_{Isl} = 4.84\,$mJy compared to our Apertif reference sample, which has $TP = 14.8\,$mJy and $PI_{Isl} = 0.78\,$mJy. 
This can also be seen in the histogram of polarised and total intensity flux density in Fig. \ref{fig:ap:PI_TP_hist}.


We note that, in contrast to Fig. \ref{fig:ap:FP_PDF}, these plots do not show the density function, but the histograms are just normalised by the total number of sources in each set for better visual comparability. 

The steep drop in polarised intensity for the Apertif reference sample is dominated by the sensitivity limitation in polarised intensity \citep[see][for details]{Berger2025}. 
A corresponding behaviour is not visible in the ApPolLo sample, indicating that no corresponding selection effect is significant for this sample. 

\subsubsection{Rotation Measure}$~$\\
The measured $RM$ is the sum of all rotation of the polarisation angle the signal undergoes along the LOS and contains no positional information of the rotating medium. To obtain the extragalactic component of the $RM$, we thus needed to first correct for the galactic $RM$ ($GRM$), to be left with the Residual $RM$ ($RRM$), via
\begin{align}
    RRM = RM - GRM\, .
    \label{eq:ap:RRM}
\end{align}
The $RRM$ consists of the contribution of the $RM$ from the local medium associated with the source itself or its direct vicinity and the contribution from the extragalactic medium along the LOS. 
Since $GRM$ is solely dependent on the foreground rotation due to the plasma of the Milky Way and therefore the sources' position on the sky and is often much stronger than the extragalactic component, we need to correct for this before comparing the ApPolLo sample to the Apertif reference sample, otherwise we would only see the imprint of the galactic magnetic field at the different fields' positions.
We calculate the $GRM$ from the map of \cite{Hutschenreuter2022}, by taking the median over a $1\,$deg diameter area around the source position for the ApPolLo sources as well as for the Apertif reference sample. 

\begin{figure}
    \centering
    \includegraphics[width = \linewidth]{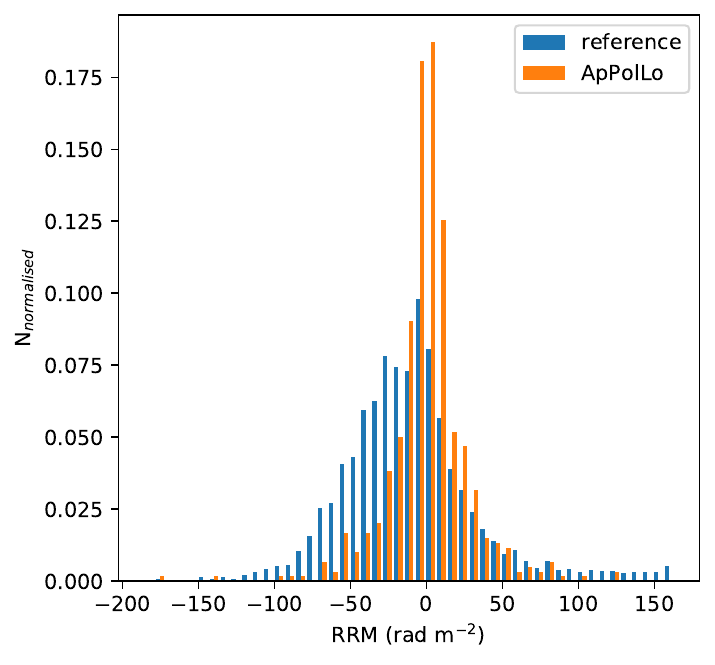}
    \caption[$RRM$ histogram of the ApPolLo and Apertif reference samples]{Histogram of the $RRM$ of the ApPolLo sample, compared to the Apertif reference sample.}
    \label{fig:ap:RRM_hist}
\end{figure}

We find the median $RRM$ of the ApPolLo sample of $0.29\,\text{rad m}^{-2}$ to be of slightly lower absolute value than the $RRM$ of the Apertif reference sample of $- 11.50\,\text{rad m}^{-2}$ (see Fig. \ref{fig:ap:RRM_hist}). However, it is more noticeable that the MAD of the ApPolLo sample is more than a factor of two smaller with $15.08\,\text{rad m}^{-2}$ compared to the MAD of $34.90\,\text{rad m}^{-2}$ for the reference sample. This indicates a significantly narrower distribution of $RRM$ of ApPolLo sources, which again can be explained with the findings by \cite{Carretti2022}, that the LOFAR RM grid catalogue sources tend to be found in low density environments. 

\subsection{Apertif PI non-detections}$~$\\
\label{sec:non-detect}
We expected polarised detections for all ApPolLo sources in our Apertif data due to the strong wavelength dependency of depolarisation effects 
\cite[see e.g. eq. \ref{eq:extrenal_depol} and][]{Sokoloff1998,Burn1966}. 
Therefore, it is interesting that we found two sources in our sample that do not show any significant polarised intensity at 1.4\,GHz. These sources were neither detected via \texttt{aperpol} nor via our peak finding strategy (see Sec. \ref{sec:peak_finding}). Visual inspection also gave no hint for any significant emission.
The non-detection of these sources may be due to sensitivity limitations or physical reasons that would cause low polarisation at short wavelengths.

Even though we did not detect the sources, we measured the flux in our Apertif $PI$ image at the LOFAR RM grid catalogue sources' positions, and calculated upper limits for the flux. These values are presented in Table \ref{tab:ap:non-det} and the measured fluxes are also shown in Fig. \ref{fig:ap:PI_TP_hist}. 
As expected in the case of non-detections due to limited sensitivity, the fluxes are on the lower end of the flux distribution.
The sensitivity scenario is also supported by the upper limit $FP$ values for both sources. The extremely high $FP_{UL}$ of source 411 suggests the sources $PI$ to be far below the sensitivity level.

We found no peculiarities that would suggest a non-detection due to physical reasons.

\begin{table}
    \centering
    \caption{ApPolLo sources not detected in Apertif.}
    \begin{tabular}{c|c|c|c|c|c}
        ID  & $\sigma_{QU}$ & $PI_{pos}$ & $FP_{pos}$ & $PI_{UL}$ & $FP_{UL}$ \\
         & $\upmu$Jy & $\upmu$Jy & $\%$ & $\upmu$Jy & $\%$ \\
        \hline \hline
        340 & 30.85         & 150.12     & 5.3\,\% & 205.00  & 7.2\,\%\\
        411 & 15.13         & 55.71      & 24.8\,\% & 100.54 & 44.8\,\%   
    \end{tabular}
    \caption*{\small\textbf{Note.} Noise and flux values for the two ApPolLo sources that are not detected in polarised intensity at 1.4\,GHz. The first column gives the source ID from the LOFAR RM grid catalogue, the second column gives the noise in the Apertif $PI$ map at the position of the LOFAR RM grid catalogue source, the third column gives the 1.4\,GHz flux measured at that exact location and column four the resulting fractional polarisation. Column five gives the flux upper limit at $95\,\%$ confidence calculated from $\sigma_{QU}$ and column six gives the respective fractional polarisation.}
    \label{tab:ap:non-det}
\end{table}


\section{Depolarisation with redshift}
\label{sec:depol}

The median fractional polarisation of our sources is $1.65\,\%$ at 144\,MHz and 
$FP_{peak}=10.39\,\%$ ($FP_{pos} = 8.62\,\% $) at 1.4\,GHz. 

We calculated the depolarisation $DP^{144MHz}_{1.4GHz}$ as 
\begin{align}
    DP = \frac{FP(\lambda_2)}{FP(\lambda_1)},
    \label{eq:depolarisation}
\end{align}
where $\lambda_1 = 21\,$cm and $\lambda_2 = 2\,$m. 
Since the fractional polarisation at metre wavelengths is usually much lower than at shorter wavelengths due to the strong wavelength dependency of depolarisation effects, we expected most sources to show $DP^{144MHz}_{1.4GHz} < 1$. 

\begin{figure}
    \centering
    \includegraphics[width = \linewidth]{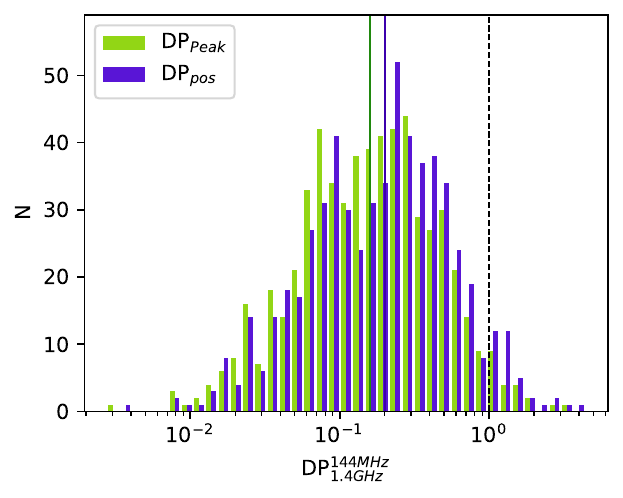}
    \caption[Histogram of the Depolarisation of ApPolLo sources]{Histogram of the depolarisation of ApPolLo sources. The dashed black line indicates $DP^{144MHz}_{1.4GHz}$ = 1, where the source's $FP$ is equal at 1.4\,GHz and 144\,MHz, the thin green line highlights the median $DP_{Peak}$ and the thin purple line the median $DP_{pos}$.}
    \label{fig:ap:depol_hist}
\end{figure}

Figure \ref{fig:ap:depol_hist} shows the histogram of the depolarisation. We find a median $DP^{144MHz}_{1.4GHz,\ \ peak} = 0.16 \pm 0.10$ ($DP^{144MHz}_{1.4GHz,\ \ pos} = 0.20 \pm 0.13$), which is in agreement with the findings of \cite{Stuardi2020} who found a median $DP^{144MHz}_{1.4GHz}$ of 0.2 for giant radio galaxies using LOFAR and the NVSS. 
21 sources, 3.5$\,\%$ of the sample, show stronger fractional polarisation at 144\,MHz than at 1.4\,GHz, using $FP_{Peak}$ as a measure. 
Using $FP_{pos}$ we find 36 (6.0$\,\%$ of the sample) sources with $DP^{144MHz}_{1.4GHz} > 1$, one of which is the Apertif non-detection source 411 (see Sec. \ref{sec:non-detect}. As explained in Sec. \ref{sec:peak_finding}, the peak positions at 1.4\,GHz and 144\,MHz do not need to align, therefore the higher number of repolarised sources, when using $FP_{pos}$ compared to using $FP_{Peak}$ is most likely just a result of this misalignment. 
In total, there are 39 sources (6.5$\,\%$ of the sample) in which repolarisation was observed in any of the two measurements.
These sources will be further analysed in a follow-up work.

Using the information from the LoTSS value added catalogue, we have redshift information for 449 ApPolLo sources and find the median redshift to be 0.59. In agreement with \cite{Carretti2022} we find only a few sources at high redshift. 
We limited our sample for all the following investigations to sources with $0 \leq z \leq 1.5$, to avoid low number statistics. After excluding also the two Apertif non-detections (Sec. \ref{sec:non-detect}), we were left with 429 sources.

\begin{figure}
    \centering
    \includegraphics[width = \linewidth]{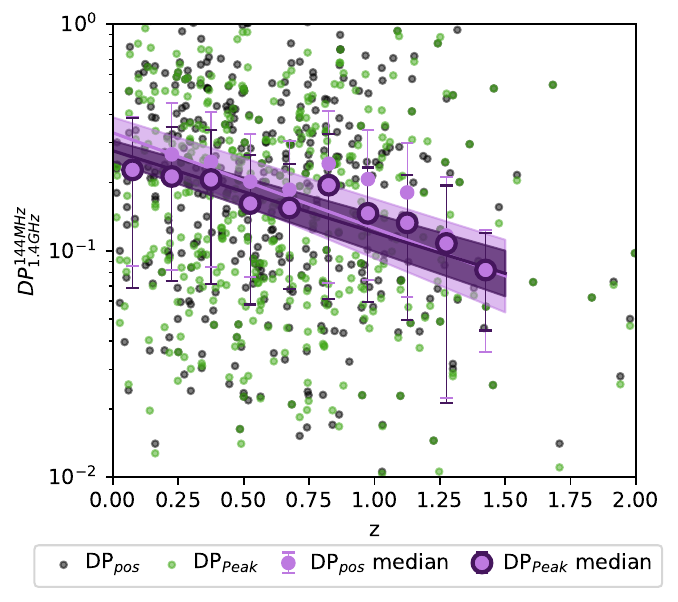}
    \caption[ApPolLo depolarisation as a function of redshift]{Depolarisation between 1.4\,GHz and 144\,MHz as a function of redshift. The small black ($DP_{pos}$ and green ($DP_{Peak}$) points show the individual sources, while the larger lilac ones represent the median in redshift bins. The $DP_{Peak}$ median values have dark edges. The errors shown are the MAD in the bin and are used to weight the data for fitting. The light lilac line is the best linear fit using the median values of $DP_{pos}$, the dark lilac line is the best fit using $DP_{Peak}$.}
    \label{fig:ap:DP_z}
\end{figure}

Assuming at least a part of the depolarisation to happen as external Faraday depolarisation from cosmic web filaments, $DP^{144MHz}_{1.4GHz}$ is expected to decrease with redshift \citep{Carretti2022}. This is due to the increasing probability for intervening filaments with increasing redshift.
In Fig. \ref{fig:ap:DP_z} we show $DP$ as a function of redshift for both measures of $FP$. 
Our data shows a decrease in Fig. \ref{fig:ap:DP_z} to be approximately linear in a log-linear plot with a Spearman correlation coefficient of $-0.96$ ($-0.80$) and a p-value of $7.3\,10^{-6}$ ($0.004$) for using $DP_{peak}$ ($DP_{pos}$). 
Fitting $DP(z) = 10^{m z + b}$ gave $m_{Peak} = -0.36 \pm 0.04$ ($m_{pos} = -0.42 \pm 0.06$) and $b_{Peak} = -0.56 \pm 0.04$ ($b_{pos} = -0.48 \pm 0.07$ as the best fit values. The best fit and its one-sigma uncertainty are also shown for both measurements of $DP$ in Fig. \ref{fig:ap:DP_z} as the dark (light) purple line and shaded area. We note that both fits agree within their uncertainties.
Since neither the median (see Fig. \ref{fig:ap:depol_hist}) nor the redshift trend of $DP$ differ significantly based on the measurement method, we will use the values measured at the peak position in the Apertif $PI$ maps (denoted with index $Peak$) for the subsequent analysis. 

\subsection{Fractional Polarisation}$~$\\
\label{sec:Fractional_polarisation}
\cite{Berger2021} found an anti-correlation between the fractional polarisation at 21\,cm and the sources redshift, that was previously not observed at this wavelength.
At metre wavelengths, this trend was later observed by \citet{Carretti2022,Carretti2023,Carretti2025}. 
The aforementioned authors conducted a study
on the basis of data from LOFAR and the NVSS and found the fractional polarisation at 144\,MHz to drop with redshift over the entire redshift range probed ($z \leq 2$), while for the fractional polarisation at 21\,cm they found this trend only to be present up to a redshift of $z \sim 0.7$, while $FP$ is constant for higher redshift.

\begin{figure}[hbtp]
    \centering
    \includegraphics[width = \linewidth]{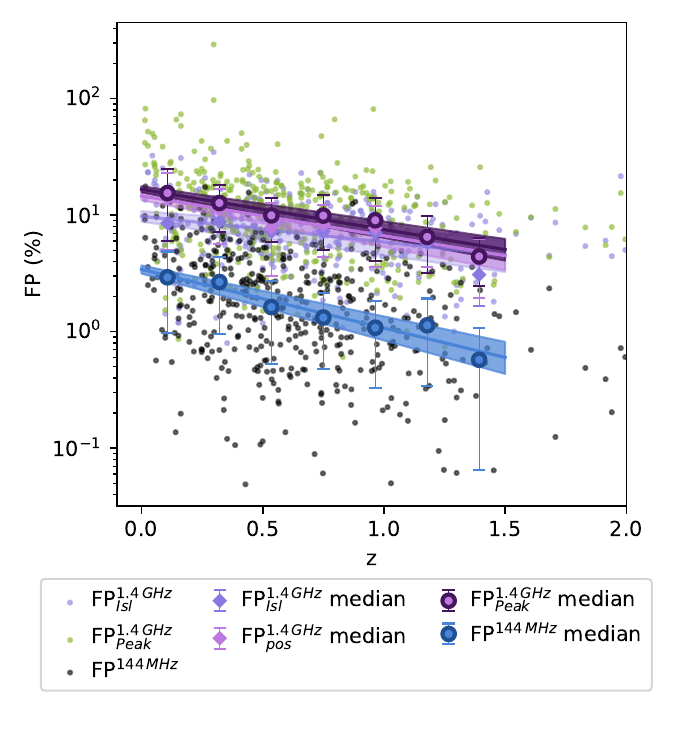}
    \caption[ApPolLo fractional polarisation as a function of redshift]{Fractional polarisation as a function of redshift for the ApPolLo sample. The small points represent the individual sources at 1.4\,GHz and 144\,MHz, respectively. The large circles and diamonds give the median binned over redshift. The best fits are shown as lines in the respective colour with the shaded area marking their $1\sigma$ uncertainties.}
    \label{fig:ap:FP_z}
\end{figure}

In Fig. \ref{fig:ap:FP_z} we present the fractional polarisation as a function of redshift for our 429 sources with $z \leq 1.5$, we used the redshift from the LOFAR DR2 RM grid catalogue for this. 
In agreement with \cite{Berger2021} and \cite{Carretti2022}, we found a decrease of median fractional polarisation with increasing redshift.
While \cite{Carretti2022} found the decrease of $FP$ for 1.4\,GHz, to flatten at around $z \sim 0.7$, the decrease remains visible in our data over our entire redshift range in Fig. \ref{fig:ap:FP_z}. 

We show all three measurements we have for $FP$ at 1.4\,GHz. Because the $PF_{Peak}$ and $FP_{pos}$ values are quite similar, we only show the individual values of $FP_{Peak}$. However, the binned values, the best fit to $FP(z) = 10^{mz+b}$ and its $1\,\sigma$ uncertainty are shown for each measure. 
The best fit parameters are presented in Tab. \ref{tab:FP_z_fit} for all three 1.4\,GHz measurements as well as for the 144\,MHz data and for the $DP(z)$ dependency, where we basically assumed the same function (see Sec. \ref{sec:depol})

\begin{table}[]
    \centering
    \caption{Best fit values for FP over z.}
    \begin{tabular}{l|c|c}
         & $m$ & $b$ \\
         \hline
         \hline
        $FP_{Peak}^{1.4\,GHz}$ & $-0.35 \pm 0.04$ & $1.22 \pm 0.02$\\[4pt]
        $FP_{pos}^{1.4\,GHz}$  & $-0.34 \pm 0.05$ & $1.16 \pm 0.03$ \\[4pt]
        $FP_{Isl}^{1.4\,GHz}$  & $-0.22 \pm 0.06$ & $0.99 \pm 0.04$ \\[4pt]
        $FP^{144\,MHz}$        & $-0.50 \pm 0.06$ & $0.53 \pm 0.03$\\[4pt]
        $DP_{Peak}$            & $-0.36 \pm 0.04$ & $-0.56 \pm 0.04$\\[4pt]
        $DP_{pos}$             & $-0.42 \pm 0.06$ & $-0.48 \pm 0.07$\\
    \end{tabular}
    \label{tab:FP_z_fit}
    \caption*{\small{\textbf{Note.}} Best for for $FP(z) = 10^{m z + b}$ and $DP(z) = 10^{m z + b}$.}
\end{table}

As visible in Fig. \ref{fig:ap:FP_z} and Tab. \ref{tab:FP_z_fit} $FP_{Peak}$ and $FP_{pos}$ agree within their uncertainties. As expected $FP_{Isl}$ gives a different slope; this value is based on the integrated flux, especially for large extended sources the measurement of the integrated flux did not work perfectly. The measurement of largely extended sources, especially in polarised intensity is more complex and we did not optimise our measurement strategy for this, as we do not have corresponding measurements at 144\,MHz.
As the $FP_{Peak}$ and the $FP_{pos}$ are most comparable to the 144\,MHz data and agree with each other we continue to use the $FP_{Peak}$ data for our subsequent analysis.

\subsection{Sensitivity Limitation}
The behaviour of fractional polarisation can be significantly effected by the survey sensitivity; for instance the fractional polarisation is known to anti-correlate with the sources total intensity, as only strong polarised sources exceed the sensitivity limitation at low total intensities. To verify if our observed trends might be the result of such selection effects, we employ a simulation of $10^7$ sources: 
The sources' redshift was drawn from a uniform distribution in the range [0,2]. We searched the LoTSS DR2 for the source closest to the drawn redshift, to determine a realistic luminosity at 144\,MHz, $P_{144\,MHz}$.
The $FP_{1.4\,GHz}$ is drawn from the PDF from \cite{Berger2025}, the $FP_{144\,MHz}$ was drawn from a log-normal PDF fitted to the data of \cite{OSullivan2023}, the spectral index $\alpha$ was drawn from a normal distribution fitted to the data of \cite{Kutkin2025}, who calculated the spectral index between Apertif 1.4\,GHz and LOFAR 144\,MHz data for more than 10 000 sources. 
From those values we calculated $P_{1.4GHz}$, the total and polarised intensities at both wavelengths and the depolarisation, to compare the total simulated sample to the subsample with the sensitivity limits of $PI_{144\,MHz} = 0.6\,$mJy, $TP_{144\,MHz}=1\,$mJy, $PI_{1.4\,GHz}=90\,\upmu$Jy, applied.
While for all other properties the simulated trends agreed with our real data, the simulation did neither show a dependency of $FP$ at either wavelength, nor of $DP$ with redshift. 
We therefore have no evidence for the observed redshift dependencies to be the result of sensitivity effects.

\section{Origin of Depolarisation}
\label{sec:depol_z}

Depolarisation is an extremely degenerate process, that is usually mathematically described by a combination of multiple depolarising layers. It is likely that the emitting medium itself also causes Faraday Rotation, possibly leading to internal Faraday Dispersion or in the case of solely regular magnetic fields differential Faraday Rotation. A full model would therefore need to include internal Faraday Dispersion and multiple different layers causing external Faraday Dispersion, including the direct source environment as well as cosmic web filaments.

Since calculations based on such a model are far beyond the capabilities of our dataset and the scope of this work, throughout this section we assume a more simplistic model, where we assume the entire depolarisation to be caused purely by external Faraday Dispersion.

\subsection{Single Layer Depolarisation}
\label{sec:depol_source_local}

Assuming the depolarisation to be purely from external Faraday Dispersion, we can use Eq. \ref{eq:extrenal_depol} to calculate the total of all $\sigma_{RM}$ along the whole LOS for each individual redshift bin in Fig. \ref{fig:ap:FP_z}.
In the absence of a better tracer for $FP_0$ we assume $FP_0 = FP(z=0)$ using our best fit from Sec. \ref{sec:Fractional_polarisation}.

Due to the wavelength dependency it is crucial where we assume the depolarisation to happen, as the wavelength changes with redshift via:
\begin{equation}
    z = \frac{\lambda_{obs} - \lambda_0}{\lambda_0}.
    \label{eq:redshift_lambda}
\end{equation}
Here, we assumed the two most extreme and simple cases where all depolarisation happens in a single depolarising layer located (1) close to the observer, so the wavelength is the detected wavelength, and (2) close to the source, so the wavelength is the emitted wavelength calculated by Eq. \ref{eq:redshift_lambda} using the central redshift of the considered redshift bin. Using this simplification we attribute everything where the effective wavelength is the emitted wavelength as 'local to the source', physically speaking this may include environmental effects within the source itself or its close environment as well as their halos or cluster environment.

\begin{figure}[htbp]
    \centering
    \includegraphics[width = \linewidth]{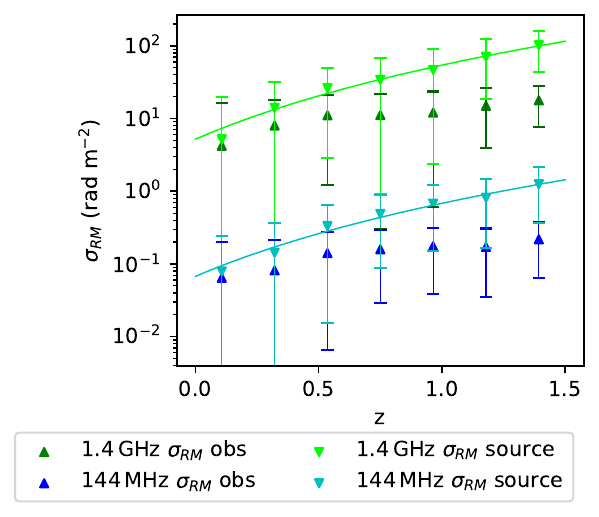}
    \caption[$\sigma_{RM}$ for different redshift ranges in Apertif and LOFAR]{Calculated $\sigma_{RM}$ for the different redshift bins used in Fig. \ref{fig:ap:FP_z}. Dark green and blue are calculated without any correction applied to the redshifted signal, thus assuming the depolarisation to occur close to the observer. Light green and blue use the emitting wavelengths, thus assuming the depolarisation to occur local to the source.}
    \label{fig:ap:sigma_z}
\end{figure}

The $\sigma_{RM}$ calculated this way is shown in Fig. \ref{fig:ap:sigma_z}, 
case (1) is shown in dark colours and case (2) in light colours.
In both cases, the values obtained from the 1.4\,GHz data are about two orders of magnitude higher than those from the 144\,MHz data.

In case (1) we assume all depolarisation to happen close to the observer, this would not allow for any evolution with redshift. However, the results of case (1) can be seen as lower limits for the total $\sigma_{RM}$.

In case (2) where we assumed the depolarising layer to be local to the source, we see a significant increase of $\sigma_{RM}$ of about an order of magnitude from the lowest to the highest redshift.
This can be interpreted as an upper limit for the total $\sigma_{RM}$ of any depolarisation happening close to the emitting source.
We fit a function of the form $\sigma_{RM} = b (1+z)^{a}$ to the calculated values from case (2) resulting with the best fit parameters $a_{1.4\,GHz}=3.38 \pm 0.18$, $b_{1.4\,GHz} = 5.2 \pm 0.7$ and $a_{144\,MHz}=3.34 \pm 0.22$, $b_{144\,MHz} = 0.068 \pm 0.010$ for the 1.4\,GHz and 144\,MHz data respectively.

The $\sigma_{RM}$ resulting from the 144\,MHz data, would not cause any detectable depolarisation at centimetre- or higher wavelengths ($DP \geq 99\,\%$), while the high values resulting from 1.4\,GHz data would completely depolarise the signal detected at 144\,MHz, even when accounting for the shorter emitting wavelengths. 

This confirms the findings of \cite{Carretti2022}, who also argued that the measured depolarisation at 1.4\,GHz would cause complete depolarisation at 144\,MHz.
Therefore, these authors concluded that the depolarisation at 1.4\,GHz must occur local to the source, while the depolarisation observed at 144\,MHz must take place further out.
This also implies that the signal measured at 144\,GHz is emitted further out from the central part of the sources, compared to the 1.4\,GHz emission, as it would otherwise travel through the same depolarising medium and be completely depolarised.
However, those results do not allow for a more detailed differentiation between local and non-local depolarisation. While we argue that the depolarisation observed at 1.4\,GHz is dominated by local effects, the depolarisation observed at 144\,MHz can occur everywhere along the line of sight, with an upper limit for the local depolarisation presented in Fig. \ref{fig:ap:sigma_z}.


\subsection{Depolarisation from cosmic filaments}
\label{sec:depol_filaments}

Even though the direct source environment may play an important role, especially in the case of dense clusters, \cite{Carretti2022} found cosmic web filaments to be suited as the origin of both the $RRM$ and $FP$ measured by LOFAR. 
Based on the simplistic calculation in section \ref{sec:depol_source_local}, where we assumed one single depolarising layer at different positions, we now assume cosmic web filaments to cause the observed depolarisation.

When assuming external Faraday dispersion caused by a number of $N_f$ filaments, we can rewrite Eq. \ref{eq:extrenal_depol} as follows:
\begin{align}
    FP &= FP_0 \cdot \exp(-2 \hat{\sigma}_{RM}^2 N_f(z) \lambda^4 )
    \label{eq:external_depol_filaments}\\
    \text{with } N_f(z) &= -1.08 z^2 + 17.89 z - 0.37,
    \label{eq:N_filaments}
\end{align}
where $\hat{\sigma}_{RM}$ is the average $\sigma_{RM}$ over all filaments intersecting the LOS. 
$N_f(z)$ is the best fit from \cite{Carretti2022} who crossmatched the LoTSS RM grid catalogue with a filament catalogue and estimated the number of filaments along the LOS from this.

\cite{Carretti2022} fitted a constant $\hat{\sigma}_{RM}$ to their data, presenting the first estimate of the average $RM$ turbulence of cosmic web filaments.
We aim to extend their analysis by allowing for $\sigma_{RM}$ to evolve with redshift and include the $z$-dependency of $\lambda$ in our calculations. 
By doing so, Eq. \ref{eq:external_depol_filaments} becomes 
\begin{align}
    FP &= FP_0 \prod_{i = 1}^{N_f} \exp(-2 \sigma_{RM,i}^2 \lambda_i^4) ,
    \label{eq:ap:depol_Nf}
\end{align}
here $\sigma_{RM,i}$ is the $RM$ dispersion of a filament at redshift $z_i$ and $\lambda_i$ is the effective wavelength at said redshift. 

\begin{figure}
    \centering
    \includegraphics[width = \linewidth]{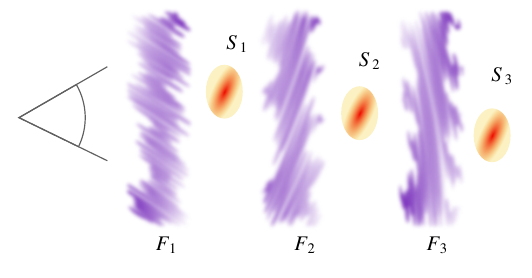}
    \caption[Sketch on depolarisation from cosmic web filaments]{Sketch to visualise the simplified example for external Faraday dispersion. $S_1,\ S_2,\ S_3$ are sources at different redshift, $F_1,\ F_2,\ F_3$ the intervening cosmic filaments.}
    \label{fig:ap:depol_example}
\end{figure}

To solve for this equation, we can first assume a simplistic case with three emitting sources $S_1,\ S_2$ and $S_3\,$ at different redshifts $z_1 < z_2 < z_3$ (see Fig. \ref{fig:ap:depol_example} for a visualisation). The LOS to $S_1$ intersects the filament $F_1$, the LOS to $S_2$ intersects the filaments $F_1$ and $F_2$, and the signal from $S_3$ passes through $F_1,\ F_2$ and $F_3$. 

The fractional polarisation measured from each source can now be calculated as follows:
\begin{align}
\label{eq:ap:depol_3sources}
\begin{split}
    FP_1 &= FP_0 \cdot \exp(-2 \sigma_{RM,1}^2 \lambda_1^4)\\
    FP_2 &= FP_0 \cdot \exp(-2 \sigma_{RM,2}^2 \lambda_2^4) \cdot \exp(-2 \sigma_{RM,1}^2 \lambda_1^4)\\
    FP_3 &= FP_0 \cdot \exp(-2 \sigma_{RM,3}^2 \lambda_3^4)\cdot \exp(-2 \sigma_{RM,2}^2 \lambda_2^4) \cdot \exp(-2 \sigma_{RM,1}^2 \lambda_1^4).\\
\end{split}
\end{align}
By dividing the above equations by one another, we can show that
\begin{equation}
\label{eq:ap:depol_3sources_simple}
\begin{split}
    \frac{FP_1}{FP_0} &= \exp(-2 \sigma_{RM,1}^2 \lambda_1^4)\\
    \frac{FP_2}{FP_1} &= \exp(-2 \sigma_{RM,2}^2 \lambda_2^4)\\
    \frac{FP_3}{FP_2} &= \exp(-2 \sigma_{RM,3}^2 \lambda_3^4).\\
\end{split}
\end{equation}
This way we can calculate $\sigma_{RM}$ of the second and third filament, even without knowledge of $FP_0$, which we, however, need to assume to be constant with $z$.

In practice, neither the exact location of the filaments is known, nor their exact number. 
Eq. \ref{eq:ap:depol_3sources_simple} can be adapted to the filaments $F_1,\ F_2$ and $F_3$ to not be single filaments, but rather an ensemble of $\Delta N_i$ statistically similar filaments, by including $\Delta N_i$ in the equation as shown in Eq. \ref{eq:external_depol_filaments}.
This way Eq. \ref{eq:ap:depol_Nf} becomes a product over $z$:
\begin{align}
    FP = FP_0 \prod_{i = z_1}^{z_{s}} \exp(-2\, \sigma_{RM,i}^2\, \Delta N_{f,i}\, \lambda_i^4)
    \label{eq:ap:depol_z}
\end{align}
where $z_s$ is the source's redshift. $\sigma_{RM,i}$ is now the average for a single filament over the considered redshift interval, $\lambda_i$ is the wavelength at the intervals mean redshift, and $\Delta N_{f,i} = N(z_n) - N(z_k)$ where $z_n$ and $z_k$ are the intervals upper and lower boundaries.

Using Eq. \ref{eq:ap:depol_z} we can now estimate $\sigma_{RM,f}(z)$ from the Apertif and LOFAR data respectively, using the best fit to $FP(z)$ to allow for a dense spacing in z without suffering from statistical variations.
Assuming both data sets to be dominated by external Faraday dispersion, we can also apply the same method to get an estimate of $\sigma_{RM,f}$ from our fit for $DP^{144MHz}_{1.4GHz}(z)$.

\begin{figure}
    \centering
    \includegraphics[width = \linewidth]{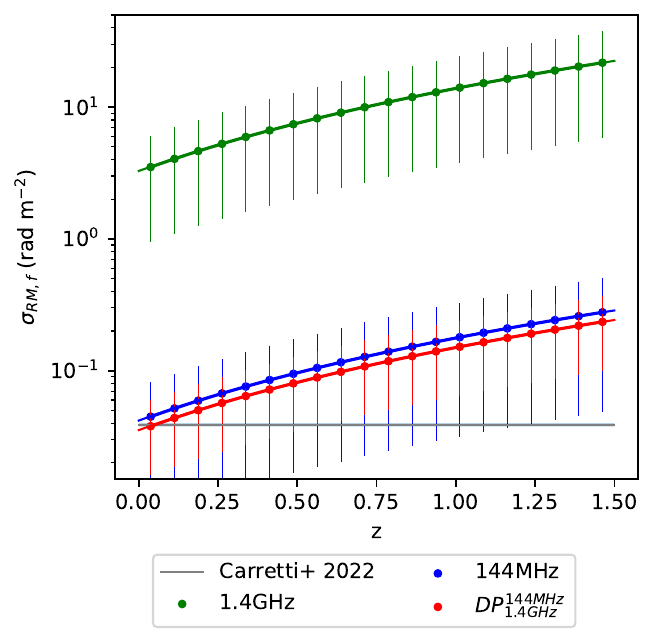}
    \caption[Redshift-dependent $\sigma_{RM}$ of cosmic web filaments]{Redshift-dependent $\sigma_{RM}$ of cosmic web filaments, as a results from the depolarisation observed at 1.4\,GHz (Apertif), 144\,MHz (LOFAR) and the depolarisation between both ($DP^{144MHz}_{1.4GHz}$). For comparison, the black line gives the constant value of $\sigma_{RM,f} = 0.0389 \pm 0.0010$ as estimated by \cite{Carretti2022}.}
    \label{fig:ap:sigma_filament_z}
\end{figure}

The resulting $\sigma_{RM,f}(z)$ is shown in Fig. \ref{fig:ap:sigma_filament_z}, together with the constant value for $\sigma_{RM,f}$ from \cite{Carretti2022}. 
The green, blue and red data points are calculated as explained above using the fits to $FP$ at 1.4\,GHz and 144MHz and the $DP^{144\,MHz}_{1.4\,GHz}$, respectively. The shown uncertainties are based on the uncertainty of the initial fits.
As expected, the $\sigma_{RM,f}$ from \cite{Carretti2022} is in agreement with our value at $z=0$.
We found $\sigma_{RM,f}$ to be best fitted by 
\begin{align}
    \sigma_{RM,f}(z) = b (z+1)^a+ c
    \label{eq:ap:sigma_fit}
\end{align}
with the best fit parameter given in Tab. \ref{tab:ap:sigma_fit_values} and also shown in Fig. \ref{fig:ap:sigma_filament_z}.

Even though we assume multiple depolarising layers, we again use the simplification that the entire depolarisation can be attributed to a single cause of depolarisation, namely cosmic web filaments. Therefore, this estimate is also to be seen as an upper limit.

\begin{table}[]
    \centering
    \caption{Best fit values for Eq. \ref{eq:ap:sigma_fit}.}
    \small
    \begin{tabular}{c|c|c|c}
                                & a & b & c \\
         \hline
         \hline
        1.4\,GHz                & $2.187 \pm 0.004$  & $2.983 \pm 0.013$ & $0.294 \pm 0.019$ \\
        144\,MHz                & $2.187 \pm 0.004$ & $0.0381 \pm 0.00017$ & $0.0038 \pm 0.00025$ \\
        $DP^{144MHz}_{1.4GHz}$  & $2.187 \pm 0.004$ & $0.0323 \pm 0.00015$ & $0.0032 \pm 0.00021$ 
    \end{tabular}
    \label{tab:ap:sigma_fit_values}
\end{table}

\section{Discussion}
\label{sec:discussion}

In Sec. \ref{sec:depol_z} we presented two possible scenarios for the observed decrease of $FP$ and $DP$ with redshift, namely depolarisation local to the source and depolarisation from cosmic web filaments.
To this end, we 
extended the calculation of $\sigma_{RM,f}$ from \cite{Carretti2022}, by allowing for a redshift-dependent $\sigma_{RM,f}$. Our estimate found $\sigma_{RM,f} \propto (z+1)^a$ with $a = 2.187 \pm 0.004$.

Figure \ref{fig:ap:sigma_filament_z} does show a large offset of about two magnitudes between the $\sigma_{RM,f}$ from the 1.4\,GHz data and the one calculated from the 144\,MHz data and from $DP^{144MHz}_{1.4GHz}$.
This offset is in agreement with our findings in Sec. \ref{sec:depol_source_local} for the more simplistic depolarising case of a single depolarising layer local to the emitting source.

As already mentioned depolarisation is a way more complex process than our model allows for.
In Sec. \ref{sec:depol_z} we assumed models with either only one layer or only one type of layer, while in reality multiple different complex layers may add to the observed depolarisation.
While such a multi layer model, that would also allow for source intrinsic evolution, is beyond the scope of this work and beyond the capacity of the present dataset, in the following we discuss how the aforementioned simplifications affect our results.

Based on Sec. \ref{sec:depol_source_local}, we concluded that the emission and depolarisation of the 1.4\,GHz signal must originate from deeper within the source, whereas the depolarisation at 144 MHz must occur further out. 
Therefore, the emission of the 144\,MHz data must also come from layers on the outskirts of the source to avoid being affected by the strong depolarising inner layers. 
This interpretation is also supported by the fact that lower-frequency radio emission is caused by older electrons, which tend to be on the outskirts of a source compared to their younger, higher-energy counterparts.
Therefore, the depolarisation observed at 1.4\,GHz is stronger effected by local (to the source) depolarisation than the 144\,MHz data.

Due to the strong $\lambda$-dependency of Faraday rotation, LOFAR data is suited to trace small magnetic field dispersions, as expected in cosmic web filaments where the magnetic fields are weak and the densities are low. 
On the other hand, the magnetic fields within the source and its direct environment are stronger, tending to lead to higher absolute values in their dispersion and are thus suited to be traced by higher frequency observations as Apertif data.
Our finding and interpretation is in agreement with similar findings of \cite{Carretti2022}. 
Those authors find external Faraday dispersion by cosmic web filaments to be the main depolarising mechanism observed at 144\,MHz.

When including the local (to the source) depolarisation, which might originate from both, internal (index $int$) and external Faraday dispersion, we need to adjust Eq. \ref{eq:ap:depol_z} by adding a factor $\xi$ for local (index $loc$) depolarisation
\begin{align}
    FP &= FP_0\, \xi_{z_s} \prod_{i = z_1}^{z_{s}} \exp(-2\, \sigma_{RM,i}\, \Delta N_{f,i}\, \lambda_i^4)\, \\
    \text{with } \xi_{z_s} &= \frac{1-\exp(-2\, \sigma_{RM,int,z_s}^2\,\lambda_{z_s}^4)}{2\, \sigma_{RM,int,z_s}^2\,\lambda_{z_s}^4} \cdot \exp(-2\, \sigma_{RM,loc,z_s}\, \lambda_{z_s}^4)
\end{align}

In the case of solely external depolarisation from cosmic filaments the assumption $FP_0 = FP(z=0)$ was reasonable, since we can assume no cosmic filament to intervene between the observer and a source at $z=0$. However, this assumption does not hold in the case of a local depolarisation component, as this depolarisation happens within or very close to the source and thus also for a source at $z=0$.
Even if not affected by non-local external Faraday dispersion, our measured (or extrapolated) $FP (z=0)$ is
\begin{align}
    FP (z&=0) = FP_0^{true} \cdot \xi_0\\
    \text{with }\xi_0 &= \frac{1-\exp(-2\, \sigma_{RM,int,0}^2\,\lambda_{z_0}^4)}{2\, \sigma_{RM,int,0}^2\,\lambda_{z_0}^4} \cdot \exp(-2\, \sigma_{RM,loc,0}\, \lambda_{z_0}^4),
\end{align}
where $FP_{0}^{true}$ is the true emitted fractional polarisation of a source at $z = 0$. 

With our approach in calculating $\sigma_{RM,f}$ in Sec. \ref{sec:depol_filaments} we propagated not the absolute effect of the local depolarisation $\xi$, but rather the ratio $\gamma^i_{i-1} = \frac{\xi_i}{\xi_{i-1}}$.
By assuming the decrease of $FP$ with $z$ to be completely due to external Faraday dispersion from cosmic filaments, we therefore do not need to assume the local component to be negligible, 
but $FP^{true}_i \cdot \gamma^{i}_{i-1}$ to be constant over the observed redshift range. 
If we assume $FP^{true}_z$ to be constant, we need $\gamma^i_{i-1} = 1$.
This would mean that the effect of the depolarisation does not change as a function of redshift. We note that $\gamma^i_{i-1} = 1$ does not imply $\sigma_{RM,loc}$ or $\sigma_{RM,int}$ to be redshift independent, but rather, due to the $\lambda^4$-dependency, an increase with $z$.
While \cite{Carretti2022} showed external Faraday dispersion to be the dominant depolarising mechanism observed in the 144\,MHz data, implying $\gamma_{i-1,\,144MHz}^i = 1$ to be a suitable assumption over the observed redshift range, it is clear from Sec. \ref{sec:depol_source_local} that this assumption is not valid for the 1.4\,GHz data. 

\begin{figure}
    \centering
    \includegraphics[width = \linewidth]{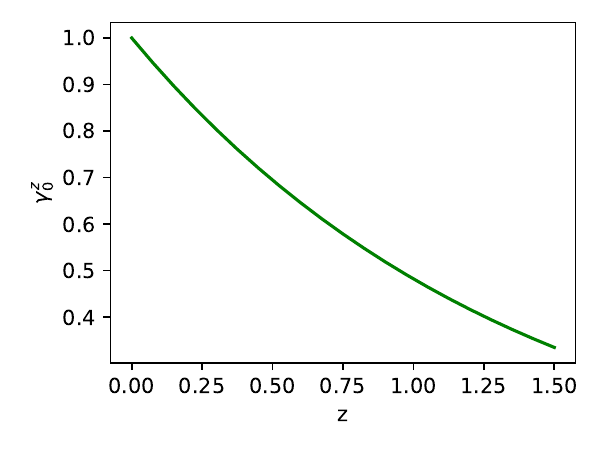}
    \caption[Relative local depolarisation at 1.4\,GHz]{$\gamma_{0,\,1.4GHz}^z$, the ratio between the effect of the depolarising component local to the source at a given redshift, to the same component at $z=0$, when assuming $\sigma_{RM,f} = 0.04 (z+1)^{2.19}+0.004$.}
    \label{fig:ap:gamma_z}
\end{figure}

Using our $\sigma_{RM,f}$ from the 144\,MHz data, we can correct our 1.4\,GHz measurements for the effect of depolarisation from cosmic filaments. The ratio of this corrected value to our measured one then yields $\gamma^z_{0,\,1.4GHz} = \frac{\xi_z}{\xi_0}$, here we do not use the ratio of $\xi$ between two neighbouring redshift bins, but between a given redshift and $z=0$. 
This is shown in Fig. \ref{fig:ap:gamma_z}. $\gamma^z_0$ decreasing by a factor of $\sim2$ between $z=0$ and $z=1.5$ corresponds to an increasing contribution of the local depolarisation with redshift. 
This is in agreement with our findings from Sec. \ref{sec:depol_source_local}, where we also find the local component's $\sigma_{RM}$ to increase as a function of redshift. 
It is worth noting that this does not directly imply an increase in the magnetic fields' strength, but can also be due to variations in other properties like the electron density or the magnetic fields' morphology (see Eq. \ref{eq:sigma_square}).

It can be argued whether the assumption of $\gamma_{0,\,144MHz}^z \sim 1$ over the observed redshift range is valid at all; because of the 1.4\,GHz data showing a strong redshift-dependence of the effect of local depolarisation (Fig. \ref{fig:ap:gamma_z}), it is reasonable to 
assume a corresponding dependency to be present in the 144\,MHz data as well, which is why our estimate for $\sigma_{RM,f}$ is to be seen as an upper limit.

When propagating the effect of depolarisation from a local component through our approach for $DP^{144MHz}_{1.4GHz}$, it can be shown that
\begin{align}
    \label{eq:ap:depol_internal}
    \frac{DP_{i}}{DP_{i-1}} &= \beta_i \,\beta_{i-1}^{-1} \cdot \exp\left(-2\, \sigma_{RM,i}\, \Delta N_{f,i}\, (\lambda_{L,z_i}^4 - \lambda_{A,z_i}^4)\right)\\
    \text{with } \beta_i &= \frac{\xi_{i,\,144MHz}}{\xi_{i,\,1.4GHz}}.\\
    \text{Assuming } \beta &\approx \beta_i  \approx \beta_{i-1}\\
    \Rightarrow \frac{DP_{i}}{DP_{i-1}} & = \exp\left(-2\, \sigma_{RM,i}\, \Delta N_{f,i}\, (\lambda_{L,z_i}^4 - \lambda_{A,z_i}^4)\right),
\end{align}
where 
$FP^{true}_z$ for both observed wavelengths is cancelled out.
In this case, the assumption $\gamma = 1$ is no longer required to hold for either wavelength. 

It is worth noting that we did not assume $FP_{0,true}^{144MHz} = FP_{0,true}^{1.4GHz}$, as this assumption would be in conflict with the previous conclusion of the 1.4\,GHz data showing strong local depolarisation, that is not visible in the 144\,MHz data, and thus being emitted from more central regions of the source.

We find that both assumptions ($\gamma_{i-1,\,144MHz}^i \sim 1$ for the calculation based on the 144\,MHz data and $\beta = const.$ for the calculation based on $DP_{1.4GHz}^{144MHz}$) to result in comparable $\sigma_{RM,f}$ values that are within each others uncertainties, as visible in Fig. \ref{fig:ap:sigma_filament_z}. 
This can be explained by $\beta \gg 1$ or in other words $\xi_{144MHz} \gg \xi_{1.4GHz}$ (note that smaller values of $\xi$ refer to stronger local depolarisation), so the relative change of the effect of the local component with redshift on the 144\,MHz data becomes negligible, therefore $\gamma_{144MHz} \sim 1$ still holds as an approximation over our observed redshift range.

With our best fit of $\sigma_{RM,f} \propto (z+1)^a$ with $a = 2.187 \pm 0.004$, and using Eq. \ref{eq:sigma_square}, we can now estimate a possible redshift dependency of $B_{f,turb} \propto (z+1)^\alpha$.
Assuming $n_e \propto (z+1)^3$ \citep{OSullivan2020,Pshirkov2016,Blasi1999} and $L \propto (z+1)^{-1.4}$ \citep{Carretti2023}, we find $\alpha \sim 0.11$, which is in agreement with the finding of $\alpha = 0.2 \pm 0.5$ from \cite{Carretti2023} for $B_f$, neither $B_f$ nor its turbulent component show significant evolution with redshift.
However, in a later study \citet{Carretti2025} revised their source selection, avoiding remaining contributions from the galactic foreground to the RRM, leading to a steeper RRM with redshift and therefore the aforementioned work found the magnetic field in cosmic web filaments to evolve with redshift. 
A detailed analysis of how remaining foreground contamination might affect our results, is beyond the scope of this work and will be discussed in a future paper.


\section{Summary and Outlook}
\label{sec:summary}

Starting with the LoTSS DR2 RM-grid catalogue \citep{OSullivan2023}, we generated a catalogue of 598 polarised components with Apertif and LOFAR polarisation data available. 

In order to compare the ApPolLo sample with a typical sample at 1.4\,GHz, we use the Apertif GOODS-N \citep{Berger2025} and SVC \citep{Adebahr2022b} samples as a reference. 
The ApPolLo sources are found to be on average more than four times stronger in total intensity, and about six times stronger in polarised intensity, compared to the Apertif reference sample. 
This also agrees with the median fractional polarisation of ApPolLo sources being significantly higher ($7.37\,\%$) than found in the reference sample ($4.48\,\%$).
Together with the finding of ApPolLo sources to show a lower $RRM$ and narrower $RRM$-spread, this is in agreement with the findings of \cite{Carretti2022} of LOFAR RM grid catalogue sources preferring low density environments.

We find the fractional polarisation observed at 1.4\,GHz, to decrease with redshift over our entire observed redshift range of $0 \leq z \leq 1.5$\,. 
A similar trend at 144\,MHz as well as for the depolarisation between both wavelengths is also visible in our data.

To investigate the origin of this depolarisation, we considered two simplified models; the first, where the entire depolarisation happens in a single layer via external Faraday Dispersion; and second, the entire depolarisation being caused by cosmic web filaments.
From the comparison of both models, we draw the following inferences:
\begin{itemize}
    \item We find the depolarisation observed in the 1.4\,GHz data to be dominated by depolarisation local to the source, with higher redshift sources showing stronger local depolarisation.
    \item For the first time we were able to present an upper limit estimate of the average redshift dependent $RM$ turbulence of filaments $\sigma_{RM,f}$. 
    This calculation is based on the 144\,MHz data, the 1.4\,GHz data and the depolarisation between both. In the first two cases we assume the effect of depolarisation local to the source $\xi_i$ on the individual data sets to be constant with $z$ ($\gamma^i_{i-1} = \frac{\xi_i}{\xi_{i-1}} = 1$), while in the latter case we assume the effect to develop equally at both observational wavelengths with redshift ($\beta =\frac{\xi_{i,\,144MHz}}{\xi_{i,\,1.4GHz}} = const.$).
    We find a redshift dependency of $\sigma_{RM,f} = b \cdot (z+1)^{2.19} + c$ for all three approaches. It is worth noting that $b$ is rather small when based on the 144\,MHz data and $DP^{144MHz}_{1.4GHz}$, since LOFAR is only sensitive to small $\sigma_{RM}$.
    \item From the comparison of $\sigma_{RM,f}$ calculated from the 144\,MHz data to that calculated from $DP^{144MHz}_{1.4GHz}$, we find that the local component of the depolarisation is much weaker for the 144\,MHz data than for the 1.4\,GHz data. Therefore, the 144\,MHz emission occurs on the outskirts of the source.
    \item Our findings of $\sigma_{RM,f} \propto (z+1)^{2.19}$ correspond to no significant evolution of the turbulent part of the physical magnetic field in cosmic web filaments with redshift.
\end{itemize}

Our analysis complements previous $RM$ based investigation of magnetic fields in cosmic web filaments, by presenting a novel method to use the fractional polarisation in order to investigate the turbulent magnetic field.

This investigation is a pilot study, using only parts of the entire AWES survey area, highlighting the scientific use of Apertif data, especially in synergy with LOFAR. Future analysis of the full AWES will allow for more detailed studies and improve the statistical uncertainties. 
Polarisation analysis of AMES will allow us to probe fainter polarised sources, possibly probing higher redshifts and also get higher S/N for sources comparable to those in this work.
A more detailed analysis of the depolarisation occurring local to the source may be possible by techniques such as $QU$-fitting, where we are limited to sources with a sufficiently high S/N and frequency coverage. This kind of analysis would also benefit from utilising data in the wavelength range between Apertif and LOFAR, where the upcoming SKA will be of great benefit. 

A full polarisation catalogue of AWES will cover approximately 2200\,$deg^2$ \cite{Adams2022} with a source density of $~23/deg^2$ \cite{Berger2025} with AMES achieving even higher source density, though on smaller area. This will yield an important northern sky counterpart to ASKAP and MeerKAT surveys, allowing for best synergy with LOFAR.



\begin{acknowledgements}

AB acknowledges financial support by the German Federal Ministry of Research, Technology and Space (BMFTR) under ErUM-Pro grant 05A23PBA (Verbundprojekt D-MeerKAT III). 
The research at RUB was supported by funding from the German Research Foundation DFG, within the Collaborative Research Center SFB1491 "Cosmic Interacting Matters - From Source to Signal".

\end{acknowledgements}

%
  \bibliographystyle{aa} 
  \bibliography{bibliothek} 
%





\end{document}